\documentclass[aps,prl,twocolumn,groupedaddress]{revtex4}

\usepackage{epsfig}
\usepackage{dcolumn}
\usepackage{bm}

\begin{document}

\def\d{{\rm d}}
\def\md{m_{\tilde{\chi}_1^0}}
\def\mn{m_{\tilde{\chi}_2^0}}
\def\mc{m_{\tilde{\chi}_1^\pm}}
\def\ms{m_{\tilde{q}}}
\def\ml{m_{\tilde{l}}}
\def\lp{\left. }
\def\rp{\right. }
\def\lr{\left( }
\def\rr{\right) }
\def\le{\left[ }
\def\re{\right] }
\def\lg{\left\{ }
\def\rg{\right\} }
\def\lb{\left| }
\def\rb{\right| }
\def\beq{\begin{equation}}
\def\eeq{\end{equation}}
\def\bea{\begin{eqnarray}}
\def\eea{\end{eqnarray}}

\preprint{IPHC-PHENO-09-02}
\preprint{LPSC 09-087}
\title{Transverse-momentum resummation for gaugino-pair production at hadron colliders}
\author{Jonathan Debove$^a$}
\author{Benjamin Fuks$^b$}
\author{Michael Klasen$^a$}
\email[]{klasen@lpsc.in2p3.fr}
\affiliation{$^a$ Laboratoire de Physique Subatomique et de Cosmologie,
 Universit\'e Joseph Fourier/CNRS-IN2P3/INPG,
 53 Avenue des Martyrs, F-38026 Grenoble, France \\
 $^b$ Institut Pluridisciplinaire Hubert Curien/D\'epartement Recherche Subatomiques,
 Universit\'e de Strasbourg/CNRS-IN2P3,
 23 Rue du Loess, F-67037 Strasbourg, France}
\date{\today}
\begin{abstract}
We present a first precision analysis of the transverse-momentum spectrum of
gaugino pairs produced at the Tevatron and the LHC with center-of-mass energies
of 1.96 and 10 or 14 TeV, respectively. Our calculation is based on a universal
resummation formalism at next-to-leading logarithmic accuracy, which is
consistently
matched to the perturbative prediction at ${\cal O}(\alpha_s)$. Numerical results
are given for the ``gold-plated'' associated production of neutralinos and
charginos decaying into three charged leptons with missing transverse energy as
well as for the pair production of neutralinos and charginos at two typical
benchmark points in the constrained MSSM. We
show that the matched resummation results differ considerably from the Monte
Carlo predictions employed traditionally in experimental analyses and discuss the
impact on the determination of SUSY mass parameters from derived transverse-mass
spectra. We also investigate in detail the theoretical uncertainties coming from
scale and parton-density function variations and non-perturbative effects.
\end{abstract}
\pacs{12.38.Cy,12.60.Jv,13.85.Qk,14.80.Ly}
\maketitle


\vspace*{-100mm}
\noindent IPHC-PHENO-09-02\\
\noindent LPSC 09-087\\
\vspace*{77mm}

\section{Introduction}
\label{sec:1}

Weak-scale supersymmetry (SUSY) is a very well motivated extension of the
Standard Model (SM) of particle physics. It can break the electroweak symmetry
radiatively, allows for its grand unification with the local gauge symmetry of
strong interactions, offers a natural explanation of the large hierarchy between
electroweak and gravitational interactions, and appears naturally in string
theories \cite{Nilles:1983ge}. Among the new particles predicted by the Minimal
Supersymmetric SM (MSSM), the fermionic partners of the neutral and charged gauge
and Higgs bosons, called neutralinos and charginos, may be relatively light, and
the lightest neutralino, stabilized by an at least approximate $R$-symmetry,
represents one of the most promising dark matter candidates, whose
gaugino/higgsino decomposition has important consequences for cosmology. The
search for SUSY particles and the identification of their properties have thus
become defining tasks of the current hadron collider program. Particular
attention has since long been paid to the production of gauginos
\cite{Barger:1983wc}, which are produced either directly or through squark/gluino
decays, decay themselves leptonically and are easily identifiable
at the Tevatron \cite{Aaltonen:2008pv} and at the LHC \cite{Aad:2009wy}.
The Tevatron collaborations CDF and D0 have already published several SUSY
searches in the ``gold-plated'' trilepton channel \cite{Aaltonen:2008pv}. For the
LHC, the CMS collaboration estimate the reach of the direct channel to be
relatively modest (universal gaugino mass parameter $m_{1/2}<180$ GeV) and
relevant only for scenarios with heavy colored particles (universal scalar masses
$m_0>1000$ GeV). However, the ATLAS collaboration have shown that a discovery can
be made in this channel for a wider range of gaugino masses ($m_{1/2}\leq 360$
GeV) and independently of $m_0$ already with a low luminosity of a few fb$^{-1}$
\cite{Aad:2009wy}. Gaugino pair production is therefore a very important SUSY
discovery channel at both currently running hadron colliders.




For an efficient suppression of the SM background from vector-boson and top-quark
production and a precise determination of the underlying SUSY-breaking model and
masses, an accurate theoretical calculation of the signal (and background) cross
section is imperative. As the lightest SUSY particle (LSP) escapes undetected, the
key distribution for SUSY discovery and measurements is the missing
transverse-energy
($\not{\!\!E}_T$) spectrum, which is typically restricted by a cut of 20
GeV at the Tevatron and 30 GeV at the LHC. While the SUSY particle pair is
produced with zero transverse momentum ($p_T$) in the Born approximation, the
possible radiation of gluons from the quark-antiquark initial state or the
splitting of gluons into quark-antiquark pairs at ${\cal O}(\alpha_s)$ in the
strong coupling constant induces transverse momenta extending to quite
substantial values and must therefore be taken into account
\cite{Beenakker:1996ch}. In addition, the perturbative calculation diverges at
small $p_T$, indicating the need for a resummation of soft-gluon radiation to all
orders \cite{Bozzi:2006fw}. Only after a consistent matching of the perturbative
and resummed calculations an accurate description of the (missing) transverse
energy spectrum and precise measurements of the SUSY particle masses can be
achieved.

In this Letter, we report on the first precision analysis of the
transverse-momentum spectrum of gaugino pairs produced at the Tevatron and the LHC
with center-of-mass energies of 1.96 and 10 or 14 TeV, respectively. We briefly
describe in the following section our implementation of the resummation formalism,
which has been improved with respect to the original proposal in numerous
respects, and present then numerical results for the production of various gaugino
pairs at two typical MSSM benchmark points. We also discuss the impact of the
computed precise transverse-momentum spectrum on the determination of SUSY mass
parameters and investigate in detail the remaining theoretical uncertainties
coming from scale and parton-density function variations and non-perturbative
effects.

\section{Transverse-momentum resummation}
\label{sec:2}

In the Born approximation, the production of neutralinos and charginos at hadron
colliders
\bea
 p\bar{p},~pp \to q\,\bar{q}^\prime +X \to \tilde{\chi}_i\, \tilde{\chi}_j +X
\eea
is induced by the quarks $q$ and antiquarks $\bar{q}'$ in the initial
(anti-)protons and is mediated by $s$-channel electroweak gauge-boson and $t$- and
$u$-channel squark exchanges. Its partonic cross section $\sigma^{(0)}$ can be
expressed in terms of the gaugino and squark masses
$m_{\tilde{\chi}^{0,\pm}_{i,j}}$ and $m_{\tilde{q}}$, the masses of the
electroweak gauge bosons, the Mandelstam variables $s$, $t$ and $u$, and
generalized charges \cite{Barger:1983wc}.

At leading order (LO) in the strong coupling constant, ${\cal O}(\alpha_s)$,
virtual loop and real parton emission corrections must be taken into account
\cite{Beenakker:1996ch}. The latter induce transverse momenta of the gaugino pair,
that extend typically to values of the order of the gaugino mass. In the
small-$p_T$ region, where the bulk of the events is produced, the convergence of
the perturbative expansion is spoiled due to the presence of large logarithms
$\alpha_s^n/p_T^2\ln^m(M^2/p_T^2)$ with $m\leq2n-1$.
These must be resummed to all orders in impact parameter ($b$) space,
\beq
 {\d\sigma^{\rm RES}\over\d p_T^2}(p_T,M,s) =
 {M^2\over s} \int_0^\infty \d b \; {b\over 2} \; J_0 (b p_T) \;
 {\cal W}(b,M,s),
\eeq
in order to correctly implement transverse-momentum conservation. Here, 
$b$ describes the minimal distance of the two incident particles in the limit of
no interaction and is the conjugate variable of the transverse momentum $p_T$;
$J_0(x)$ is the $0^{th}$-order Bessel function; $M$ is the invariant mass of
the gaugino pair; and Eq.\ (2) is evaluated numerically as in Ref.\ 
\cite{Bozzi:2005wk}.
After a Mellin transform with respect to $z=M^2/s$, the
hadronic cross section simplifies from a convolution to a product of the
parton densities evaluated at the factorization scale $\mu_F$ with the partonic
cross section, the $N$-moment of the latter
\beq
 {\cal W}_N(b,M) =
 {\cal H}_N(M) \exp\le{\cal G}_N(L)\re
\eeq
factorizing further into a $b$-independent function
\beq
 {\cal H}_N(M) =
 \sigma^{(0)}(M)\le 1+{\alpha_s\over\pi}\;{\cal H}_N^{(1)}(M/Q) + \dots\;\re,
\eeq
that is therefore finite as $p_T\to0$ or $b\to\infty$, and an exponential form
factor
\beq
 {\cal G}_N(L) = 
 Lg^{(1)}(\alpha_s L)+g^{(2)}(\alpha_s L) + 
 \dots,
\eeq
which resums the divergent leading and next-to-leading contributions in the
logarithm $L = \ln(Q^2b^2/b_0^2)$ through the process-independent functions
$g^{(1,2)}$. While $b_0 = 2e^{-\gamma_E}$ is of kinematical origin, the scale
$Q\sim M$ accounts for the arbitrary separation of the two factors. Unphysical
logarithmic divergences at $b\to0$ are regularized by replacing $L$ with
$\tilde{L} = \ln(Q^2b^2/b_0^2+1)$. The evolution from the factorization scale
$\mu_F$ and the low scale $b_0/b$ to the high scale $Q$ is performed in
${\cal H}_N (M)$ and $\exp[{\cal G}_N]$, respectively, leaving the latter
$\mu_F$-independent and allowing for a convolution of the partonic cross section
with the parton densities in the (anti-)protons at $\mu_F$ \cite{Bozzi:2005wk}.

Although SUSY particles appearing in the virtual corrections must be taken into
account for a proper renormalization of ultraviolet divergences, their heavy
masses are known to leave only little imprint numerically. In particular, the
ratios of NLO/LO total cross sections vary only from 1.26-1.28 at the
Tevatron and 1.24-1.28 at the LHC for squark masses of 350-1000 GeV (for
the similar case of sleptons, see also Fig.\ 4(a) in Beenakker, Klasen et al.\
\cite{Beenakker:1996ch}). The restoration of the
equality of the weak scalar $(\hat{g})$ and vector gauge couplings $(g)$
furthermore requires the introduction of a finite SUSY-restoring counter term,
$\hat{g}=g[1-\alpha_s/(6\pi)]$. Finally, quark-gluon initial states can induce the
production of gauginos in association with real squarks decaying into quarks and
gauginos. The singularity associated with the pole of the on-shell squark
propagator is regularized by a small finite width
$\Gamma_{\tilde{q}}\sim 10^{-2}\ms$ (the exact value has little influence
numerically), and the on-shell contribution is subtracted from the total
cross section to avoid double-counting \cite{Beenakker:1996ch,Li:2007ih}.

\section{Numerical results}
\label{sec:3}

To obtain a valid hadronic cross section at all values of $p_T$, the ${\cal O}
(\alpha_s)$ (LO) and resummed (RES) partonic cross sections are matched by
subtracting from their sum the perturbatively expanded (EXP) resummed cross
section,
\bea
 {\d\sigma\over\d p_T^2}&=&
 {\d\sigma^{\rm LO}\over\d p_T^2}+\le
 {\d\sigma^{\rm RES}\over\d p_T^2}-
 {\d\sigma^{\rm EXP}\over\d p_T^2} \re,
\eea
and by performing numerically the necessary inverse Mellin and Fourier transforms,
kinematic integrations, and parton density convolutions. The parton densities are
evaluated in the most recent parameterization by the CTEQ collaboration
{\tt CTEQ6.6M} \cite{Nadolsky:2008zw} with $\mu_F$ (and the renormalization scale
$\mu_R$) set to the average mass $\bar{m}$ of the final state particles. The SUSY
particle masses are obtained from universal parameters defined at the grand
unification scale through the renormalization group running implemented in the
computer code {\tt SUSPECT2.3} \cite{Djouadi:2002ze}.

To be specific, we choose two minimal supergravity benchmark points. The first is
the low-mass point LM0 (SU4) with universal fermion mass $m_{1/2}=160$ GeV, scalar
mass $m_0=200$ GeV, trilinear coupling $A_0=-400$ GeV, bilinear Higgs mass
parameter $\mu>0$, and ratio of Higgs vacuum expectation values $\tan\beta=10$
\cite{Aad:2009wy}. It has been been defined by the CMS (ATLAS) collaboration with
the objective of high cross sections and thus early discovery at the LHC, as the
resulting gaugino, squark and slepton masses $\mn=\mc=113$ GeV, $\md=61$ GeV,
$\ms\simeq420$ GeV, and $\ml\simeq220$ GeV lie just beyond the current Tevatron
limits. In this scenario, the lightest chargino and second-lightest neutralino
decay with 35\% and 15\% probability through virtual sleptons to the LSP and one
and two charged leptons, respectively \cite{Muhlleitner:2003vg}.

%

Our second benchmark point is the Snowmass point (and slope, SPS) 1a' with
$m_{1/2}=250$ GeV, $m_0=70$ GeV, $A_0=-300$ GeV, $\mu>0$, and $\tan\beta=10$,
which is similar to the point LM1 (SU1) defined by the CMS (ATLAS) collaboration
and, with its SUSY particle masses of $\mn=\mc=183$ GeV, $\md=98$ GeV,
$\ms\simeq550$ GeV, and $\ml=120\dots190$ GeV, is known to be compatible with all
high-energy mass bounds and low-energy precision data. Here, the
lightest chargino and second-lightest neutralino decay almost exclusively to three
charged leptons and missing transverse energy, albeit through real sleptons, which
may be experimentally reconstructed through endpoints in kinematic distributions
\cite{AguilarSaavedra:2005pw}.

\begin{figure}
 \centering
 \epsfig{file=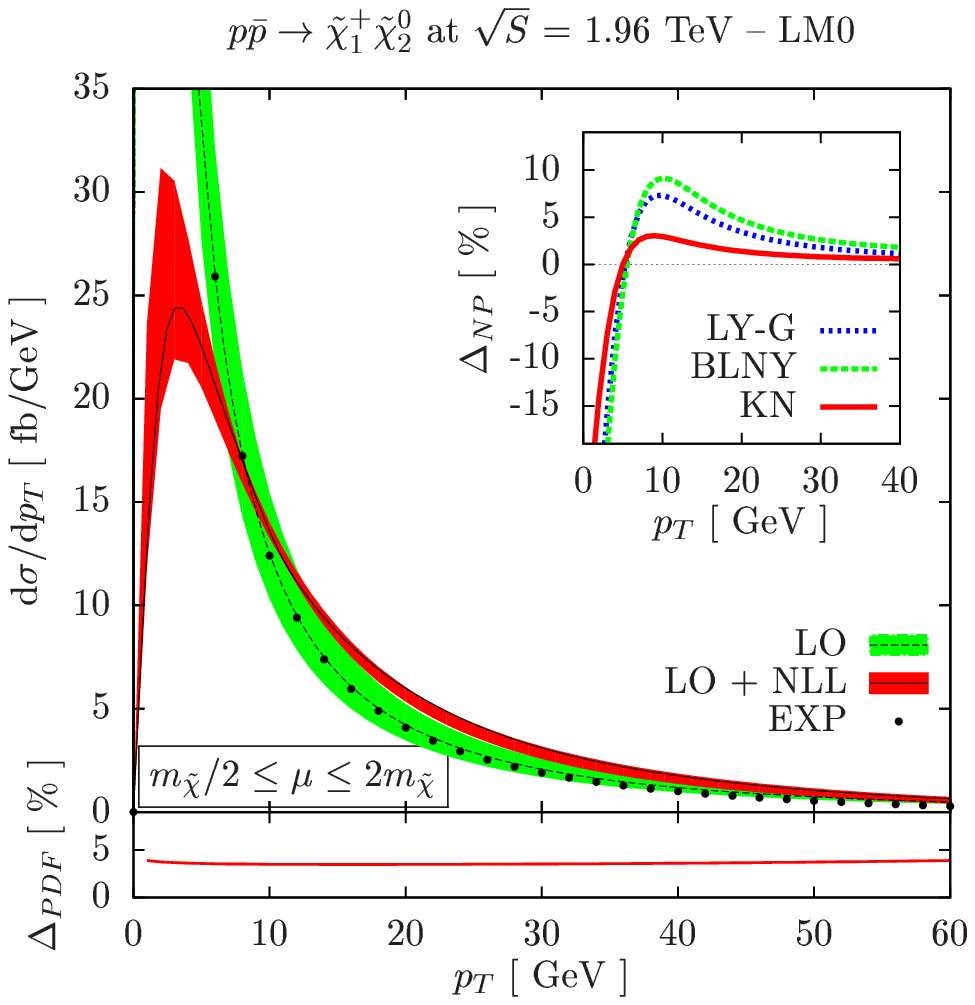,width=.86\columnwidth}
 \epsfig{file=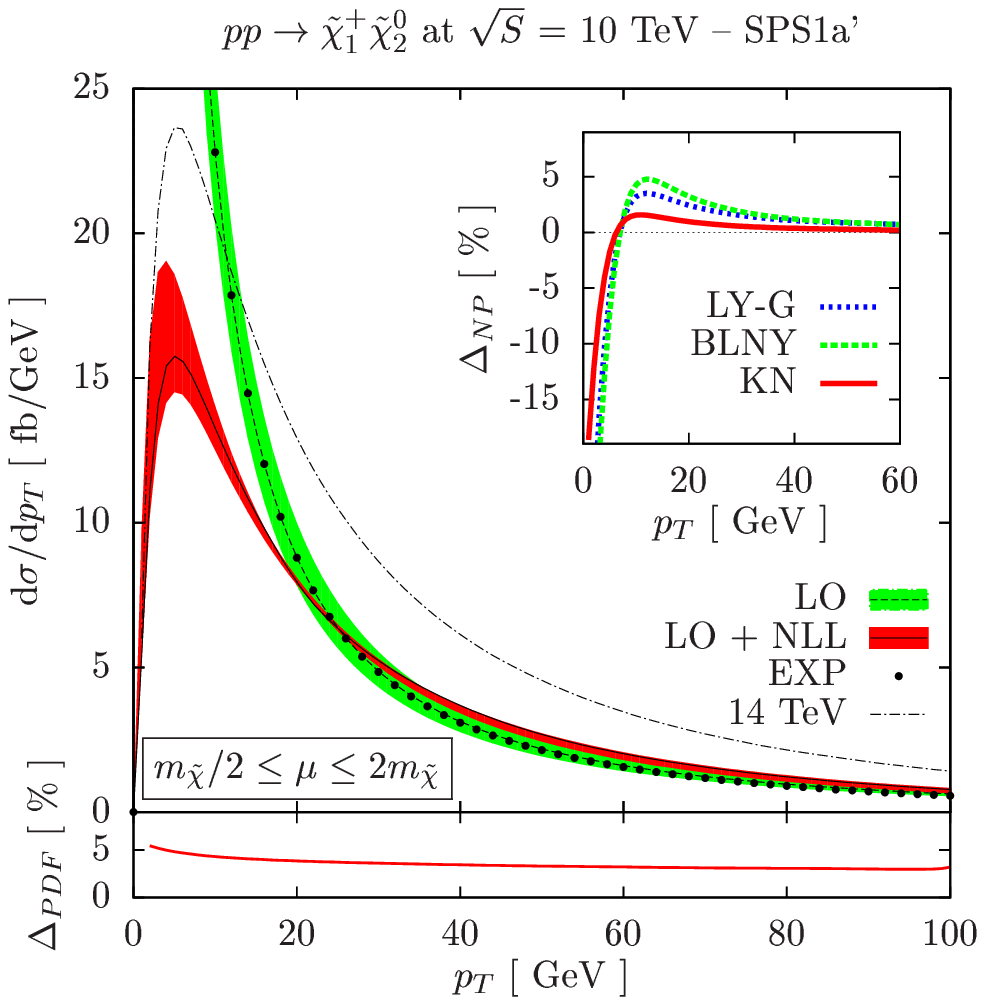,width=.86\columnwidth}
 \caption{\label{fig:1}Transverse-momentum spectra of chargino-neutralino pairs at
 the Tevatron (top) and the LHC (bottom). The LO calculation (green/dashed) is
 matched to the resummed calculation (red/full) by subtracting its fixed-order
 expansion (dotted). The scale uncertainty is shown as a shaded band, the PDF
 (below) and non-perturbative (insert) uncertainties as separate graphs, and the
 matched result for the LHC design energy of $\sqrt{S}=14$ TeV as a dot-dashed
 line (bottom).}
\end{figure}

In Fig.\ \ref{fig:1}, we show the corresponding transverse-momentum spectra of
%
%
chargino-neutralino pairs produced at run II of the Tevatron (top) and the initial
run of the LHC (bottom) with center-of-mass energies of $\sqrt{S}=1.96$ and
10 TeV,
respectively. As expected, the LO predictions (dashed curves) diverge at low
$p_T$, but become finite after matching them to the resummed predictions at
next-to-leading logarithmic (NLL) accuracy (full curves). In this region, the
perturbative expansions of the resummed predictions (dots) coincide with those at
LO, while at large $p_T$ they coincide with the resummed ones. Through
resummation, the perturbative predictions are considerably enhanced even at values
of $p_T$, which are of the order of the experimental $\not{\!\!\!E}_T$ cuts. It is
therefore important to clearly distinguish the effects induced by QCD radiation
and by the unobserved LSPs and neutrinos. By construction, the matched LO+NLL
prediction allows to reproduce the correct ${\cal O}(\alpha_s)$ correction ($K$)
factor of the total perturbative cross section after integration over $p_T$, e.g.\
of 1.26 at the Tevatron. For comparison, we also show the matched LO+NLL
$p_T$-spectrum (dot-dashed curve) for the 14 TeV design energy of the LHC, which
extends to considerably larger values of $p_T$ than at 10 TeV. The theoretical
predictions are influenced by three main sources of uncertainty: scale variations,
evaluated in the canonical range of $\mu_{F,R}/\bar{m}=0.5\dots2$ (shaded bands),
variations of the parton densities, evaluated through $\Delta_{\rm PDF}=\sqrt{
\sum_{i=1}^{22}(\d\sigma_i^+-\d\sigma_i^-)^2}/(2\d\sigma)$ along the 22
eigenvector directions defined by the CTEQ collaboration (lower curves), and three
choices of non-perturbative (NP) form factors, evaluated through $\Delta_{\rm NP}=
(\d\sigma^{\rm NP}-\d\sigma)/\d\sigma$ (inserts) \cite{Ladinsky:1993zn}. For $p_T>
5$ GeV, all theoretical uncertainties are smaller than 5\% for the LO+NLL
predictions. In particular, the 5\% PDF uncertainty is similar to the one obtained
for weak boson production \cite{Nadolsky:2008zw}.

\begin{figure}
 \centering
 \epsfig{file=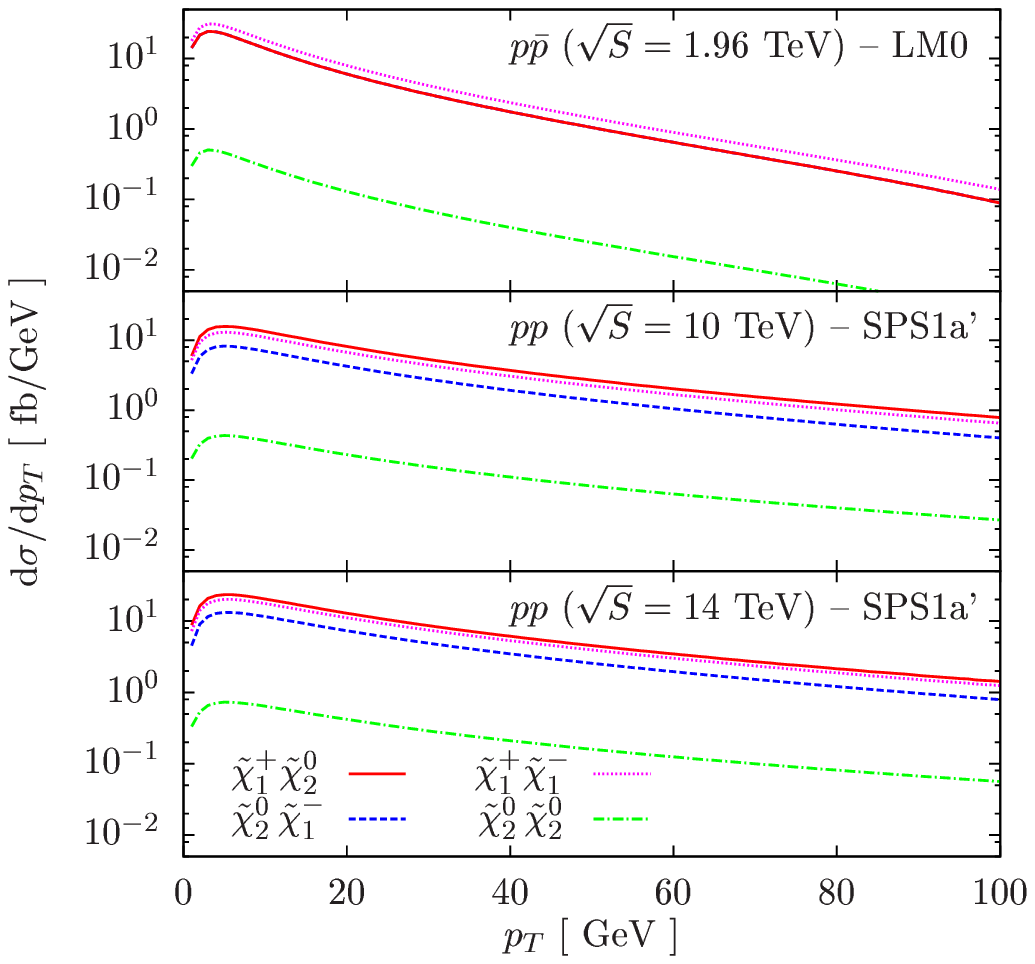,width=.86\columnwidth}
 \caption{\label{fig:2}Transverse-momentum spectra at LO+NLL for the
 associated production of charginos and neutralinos (full and dashed) as well as
 chargino (dotted) and neutralino (dot-dashed) pairs (dotted) in three different
 collider modes.}
\end{figure}

In Fig.\ \ref{fig:2}, the matched $p_T$-spectra for chargino (dotted) and
%
%
neutralino (dot-dashed) pairs are compared to those of the tri-lepton channel
(full/dashed) discussed above. While positive and negative charginos are produced
with equal rates in $p\bar{p}$ collisions at the Tevatron, their rates differ
slightly in $pp$ collisions at the LHC. The cross sections for neutralino pair
production are about one order of magnitude smaller, as the second-lightest
neutralino couples to the $s$-channel $Z^0$-boson only through its relatively
small Higgsino component.

\begin{figure}
 \centering
 \epsfig{file=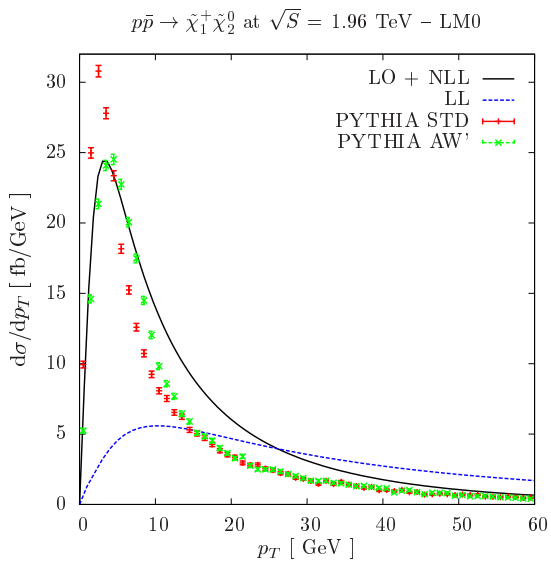,width=.86\columnwidth}
 \caption{\label{fig:3}Transverse-momentum spectra at LO+NLL (full), LL
 (dashed), and generated by the PYTHIA parton shower with default (bars) and
 tuned (crosses) parameters at the Tevatron.}
\end{figure}

In experimental analyses, QCD radiation in hadronic collisions is usually
simulated with tree-level matrix elements and parton showers based on an
exponential Sudakov form factor, which resums the leading logarithms (LL)
and some next-to-leading logarithms. In Fig.\ \ref{fig:3}, we compare therefore
%
%
our matched LO+NLL prediction (full curve) with our resummed prediction at LL
order (dashed curve) and the default (bars) and tuned (crosses) predictions of
the {\tt PYTHIA6.4} Monte Carlo (MC) generator \cite{Sjostrand:2006za}. While the
default MC prediction is clearly improved beyond the LL approximation and
approaches the LO+NLL result, it peaks at too small values of $p_T$. Tuning the
intrinsic $p_T$ of the partons in the proton to 2.1 GeV for $Z^0$-bosons (CDF tune
AW) and 4 GeV for gaugino pairs (our tune AW') improves the description of the
peak, but still underestimates the intermediate $p_T$-region and the mean value
of $p_T$ (14 GeV for {\tt PYTHIA6.4}, 15 GeV for our tune AW', and 18 GeV
for our LO+NLL prediction). This has, of course, a direct impact on the
determination of the gaugino (and slepton) masses through variables derived from
the transverse momenta of the observed leptons $p_{T,i}$ and $\not{\!\!\!E}_T$,
such as the effective mass $M_{\rm eff}=\sum_i p_{T,i}+\not{\!\!\!E}_T$
\cite{Hinchliffe:1996iu} or the stransverse mass \cite{Lester:1999tx}.
A detailed discussion of signal and background distributions in these
variables, including also the experimental resolution, is beyond the scope of this
work. Let us mention, however, that the contribution of unmeasured (low-$p_T$
or forward) or mismeasured hadronic energy to the ``fake'' $\not{\!\!E}_T$ is
under close scrutiny both at the Tevatron
and at the LHC.
The ATLAS trilepton analysis, e.g., does not identify jets with $p_T<10$
GeV, and an optional cut on jets with $p_T>20$ GeV reduces the significance
considerably. As the two LSPs are often back-to-back, the $\not{\!\!E}_T$
in the trilepton analysis is required to be relatively small ($>30$ GeV). It can
then be affected by an error of up to 10\% (Aad {\it et al.} \cite{Aad:2009wy}).

\section{Conclusion}
\label{sec:4}

In summary, we have calculated the transverse-momentum spectrum of gaugino pairs
produced at hadron colliders at next-to-leading logarithmic accuracy. We have
demonstrated that this renders the perturbative prediction finite, modifies
considerably the traditionally used Monte Carlo predictions, and reduces the
remaining theoretical uncertainties to the level of 5\%. We have also briefly
discussed the impact of our calculation on the experimental determination of the
gaugino masses, but leave a detailed experimental simulation to future work.

\acknowledgments
We thank B.\ Cl\'ement for useful discussions.
This work has been supported by a Ph.D.\ fellowship of the French ministry
for education and research and by the Theory-LHC-France initiative of the
CNRS/IN2P3.



\begin{thebibliography}{00}

\bibitem{Nilles:1983ge}
  H.~P.~Nilles,
  Phys.\ Rept.\  {\bf 110}, 1 (1984);
%
  H.~E.~Haber and G.~L.~Kane,
  Phys.\ Rept.\  {\bf 117}, 75 (1985).

\bibitem{Barger:1983wc}
  V.~Barger, R.~Robinett, W.~Keung and R.~Phillips,
  Phys.\ Lett.\  B {\bf 131}, 372 (1983);
%
  S.~Dawson, E.~Eichten and C.~Quigg,
  Phys.\ Rev.\  D {\bf 31}, 1581 (1985);
%
  G.~Bozzi, B.~Fuks, B.~Herrmann and M.~Klasen,
  Nucl.\ Phys.\  B {\bf 787}, 1 (2007);
%
  J.~Debove, B.~Fuks and M.~Klasen,
  Phys.\ Rev.\  D {\bf 78}, 074020 (2008);
%
  B.~Fuks, B.~Herrmann and M.~Klasen,
  Nucl.\ Phys.\  B {\bf 810}, 266 (2009).

\bibitem{Aaltonen:2008pv}
  T.~Aaltonen {\it et al.}  [CDF Collaboration],
  Phys.\ Rev.\ Lett.\  {\bf 101}, 251801 (2008);
%
  V.~M.~Abazov {\it et al.}  [D0 Collaboration],
  arXiv:0901.0646 [hep-ex].

%

\bibitem{Aad:2009wy}
  G.~Aad {\it et al.}  [ATLAS Collaboration],
  arXiv:0901.0512,
  in particular pp.\ 368-396 and 1643-1659;
%
  G.~Bayatian {\it et al.}  [CMS Collaboration],
  J.\ Phys.\ G {\bf 34}, 995 (2007),
  in particular Sec.\ 13.14.

\bibitem{Beenakker:1996ch}
  W.~Beenakker, R.~H\"opker, M.~Spira and P.~Zerwas,
  Nucl.\ Phys.\  B {\bf 492}, 51 (1997);
%
  W.~Beenakker, M.~Kr\"amer, T.~Plehn, M.~Spira and P.~Zerwas,
  Nucl.\ Phys.\  B {\bf 515}, 3 (1998);
%
  E.~Berger, M.~Klasen and T.~Tait,
  Phys.\ Rev.\  D {\bf 59}, 074024 (1999);
%
  W.~Beenakker, M.~Klasen, M.~Kr\"amer, T.~Plehn, M.~Spira and P.~Zerwas,
  Phys.\ Rev.\ Lett.\  {\bf 83}, 3780 (1999);
%
  E.~Berger, M.~Klasen and T.~Tait,
  Phys.\ Lett.\  B {\bf 459}, 165 (1999);
%
  E.~Berger, M.~Klasen and T.~Tait,
  Phys.\ Rev.\  D {\bf 62}, 095014 (2000).

\bibitem{Bozzi:2006fw}
  G.~Bozzi, B.~Fuks and M.~Klasen,
  Phys.\ Rev.\  D {\bf 74}, 015001 (2006);
%
  G.~Bozzi, B.~Fuks and M.~Klasen,
  Nucl.\ Phys.\  B {\bf 777}, 157 (2007);
%
  G.~Bozzi, B.~Fuks and M.~Klasen,
  Nucl.\ Phys.\  B {\bf 794}, 46 (2008);
%
  B.~Fuks, M.~Klasen, F.~Ledroit, Q.~Li and J.~Morel,
  Nucl.\ Phys.\  B {\bf 797}, 322 (2008).

\bibitem{Bozzi:2005wk}
  G.~Bozzi, S.~Catani, D.~de Florian and M.~Grazzini,
  Nucl.\ Phys.\  B {\bf 737}, 73 (2006).

\bibitem{Li:2007ih}
  C.~S.~Li, Z.~Li, R.~J.~Oakes and L.~L.~Yang,
  Phys.\ Rev.\  D {\bf 77}, 034010 (2008).

\bibitem{Nadolsky:2008zw}
  P.~M.~Nadolsky {\it et al.},
  Phys.\ Rev.\  D {\bf 78}, 013004 (2008).

\bibitem{Djouadi:2002ze}
  A.~Djouadi, J.~L.~Kneur and G.~Moultaka,
  Comput.\ Phys.\ Commun.\  {\bf 176}, 426 (2007).

\bibitem{Muhlleitner:2003vg}
  M.~M\"uhlleitner, A.~Djouadi and Y.~Mambrini,
  Comput.\ Phys.\ Commun.\  {\bf 168}, 46 (2005).

\bibitem{AguilarSaavedra:2005pw}
  J.~A.~Aguilar-Saavedra {\it et al.},
  Eur.\ Phys.\ J.\  C {\bf 46}, 43 (2006).

\bibitem{Ladinsky:1993zn}
  G.~A.~Ladinsky and C.~P.~Yuan,
  Phys.\ Rev.\  D {\bf 50}, 4239 (1994);
%
  F.~Landry, R.~Brock, P.~M.~Nadolsky and C.~P.~Yuan,
  Phys.\ Rev.\  D {\bf 67}, 073016 (2003);
%
  A.~V.~Konychev and P.~M.~Nadolsky,
  Phys.\ Lett.\  B {\bf 633}, 710 (2006).

\bibitem{Sjostrand:2006za}
  T.~Sj\"ostrand, S.~Mrenna and P.~Skands,
  JHEP {\bf 0605}, 026 (2006);
%
  R.~Field  [CDF Collaboration],
  FERMILAB-PUB-06-408-E (2006).

\bibitem{Hinchliffe:1996iu}
  I.~Hinchliffe, F.~E.~Paige, M.~D.~Shapiro, J.~Soderqvist and W.~Yao,
  Phys.\ Rev.\  D {\bf 55}, 5520 (1997).

\bibitem{Lester:1999tx}
  C.~G.~Lester and D.~J.~Summers,
  Phys.\ Lett.\  B {\bf 463}, 99 (1999).


\end{thebibliography}
\end{document}